%% file: main.tex
\documentclass[sigconf]{acmart}
\AtBeginDocument{%
  \providecommand\BibTeX{{%
    \normalfont B\kern-0.5em{\scshape i\kern-0.25em b}\kern-0.8em\TeX}}}
\copyrightyear{2024}
\acmYear{2024}
\setcopyright{acmlicensed}\acmConference[WWW '24 Companion]{Companion Proceedings of the ACM Web Conference 2024}{May 13--17, 2024}{Singapore, Singapore}
\acmBooktitle{Companion Proceedings of the ACM Web Conference 2024 (WWW '24 Companion), May 13--17, 2024, Singapore, Singapore}
\acmDOI{10.1145/3589335.3648334}
\acmISBN{979-8-4007-0172-6/24/05}

\newcommand{\baby}{\textsc{CoFARS}\xspace}
\AtBeginDocument{%
  \providecommand\BibTeX{{%
    \normalfont B\kern-0.5em{\scshape i\kern-0.25em b}\kern-0.8em\TeX}}}
\newcommand{\paratitle}[1]{\vspace{1.5ex}\noindent\textbf{#1}}

\acmPrice{15.00}
\usepackage{graphicx}
\usepackage{natbib}
\usepackage{doi}
\usepackage{multirow}
\usepackage{verbatim}
\usepackage{algorithm}
\usepackage{algorithmic} %format of the algorithm
\usepackage{xspace}
\renewcommand{\algorithmicrequire}{\textbf{INPUT:}}
\renewcommand{\algorithmicensure}{\textbf{OUTPUT:}}
\begin{document}
\begin{sloppypar}

\title{Context-based Fast Recommendation Strategy for Long User Behavior Sequence in Meituan Waimai}

% \author{Zhichao Feng$^{1,*}$, Junjie Xie$^{2,*}$, Kaiyaun Li$^{1}$, Yu Qin$^{2}$, Pengfeng Wang$^{1,\dag}$, Qianzhong Li$^{2}$, Bin Yin$^{2}$, Xiang Li$^{3,\dag}$, Wei Lin$^{3}$, Shangguang Wang$^{1}$}
% \thanks{$*$Both authors contributed equally to this work.}
% \thanks{$\dag$ Corresponding authors.}
% \affiliation{\institution{$^1$School of Computer Science, Beijing University of Posts and Telecommunications} \country{}} 
% \affiliation{\institution{$^2$Meituan} \country{}}
% \affiliation{\institution{$^3$Unaffiliated} \country{}} 

% \author{Zhichao Feng\textsuperscript{1}, JunJie Xie\textsuperscript{2}, Kaiyuan Li\textsuperscript{1}, Yu Qin\textsuperscript{2}, Pengfei Wang\textsuperscript{1}, Qianzhong Li\textsuperscript{2}, Bin Yin\textsuperscript{2}, Xiang Li\textsuperscript{2}* (Corresponding Author), Wei Lin\textsuperscript{2}, Shangguang Wang\textsuperscript{1}}
% \affiliation{%
%   \institution{\textsuperscript{1}School of Computer Science, Beijing University of Posts and Telecommunications, Beijing, China}
% }
% \affiliation{%
%   \institution{\textsuperscript{2}Meituan, Beijing, China}
% }
% \email{fengzc@bupt.edu.cn; 695799195@qq.com; tsotfsk@bupt.edu.cn; qinyu12@meituan.com; wangpengfei@bupt.edu.cn; {liqianzhong, yinbin05, xiangli, weilin}@meituan.com; sgwang@bupt.edu.cn}

\author[*]{Zhichao Feng}
\authornotemark[1]
\affiliation{
\institution{
Beijing University of Posts and Telecommunications}
\city{Beijing}
  \country{China}}
\email{fengzc@bupt.edu.cn}

\author[**]{JunJie Xie}
\authornote{Both authors contributed equally to this research.}
\affiliation{%
  \institution{Meituan}
  \city{Beijing}
  \country{China}}
\email{xiejunjie02@meituan.com}

\author[*]{Kaiyuan Li}
\affiliation{%
  \institution{
Beijing University of Posts and Telecommunications}
\city{Beijing}
  \country{China}}
\email{tsotfsk@bupt.edu.cn}

\author[**]{Yu Qin}
\affiliation{%
  \institution{Meituan}
  \city{Beijing}
  \country{China}}
  \email{qinyu12@meituan.com}

\author[*]{Pengfei Wang}
\authornotemark[2]
\affiliation{%
  \institution{
Beijing University of Posts and Telecommunications}
\city{Beijing}
  \country{China}}
  \email{wangpengfei@bupt.edu.cn}

\author[**]{Qianzhong Li\\Bin Yin}
\affiliation{%
  \institution{Meituan}
  \city{Beijing}
  \country{China}}
 \email{{liqianzhong,yinbin05}@meituan.com	}

\author[***]{Xiang Li}
% \authornotemark{2}
\authornote{Corresponding author.}
\affiliation{%
  \institution{Meituan}
  \city{Beijing}
  \country{China}}
  \email{lixiang245@meituan.com}

\author[***]{Wei Lin}
\affiliation{%
  \institution{Meituan}
  \city{Beijing}
  \country{China}}
  \email{linwei31@meituan.com}

\author{Shangguang Wang}
\affiliation{%
  \institution{Beijing University of Posts and Telecommunications}
  \city{Beijing}
  \country{China}}
  \email{sgwang@bupt.edu.cn}



\renewcommand{\shortauthors}{Zhichao Feng et.al.}

\begin{abstract}
In the recommender system of Meituan Waimai, we are dealing with ever-lengthening user behavior sequences, which pose an increasing challenge to modeling user preference effectively. A number of existing sequential recommendation models struggle to capture long-term dependencies, or they exhibit high complexity, both of which make it difficult to satisfy the unique business requirements of Meituan Waimai's recommender system.

To better model user interests, we consider selecting relevant sub-sequences from users' extensive historical behaviors based on their preferences. In this specific scenario, we've noticed that the contexts in which users interact have a significant impact on their preferences. For this purpose, we introduce a novel method called \textbf{Co}ntext-based \textbf{Fa}st \textbf{R}ecommendation \textbf{S}trategy (referred to as \baby) to tackle the issue of long sequences. We first identify contexts that share similar user preferences with the target context and then locate the corresponding Points of Interest (PoIs) based on these identified contexts. This approach eliminates the necessity to select a sub-sequence for every candidate PoI, thereby avoiding high time complexity. Specifically, we implement a prototype-based approach to pinpoint contexts that mirror similar user preferences. To amplify accuracy and interpretability, we employ Jensen–Shannon(JS) divergence of PoI attributes such as categories and prices as a measure of similarity between contexts. Subsequently, we construct a temporal graph that encompasses both prototype and context nodes to integrate temporal information. We then identify appropriate prototypes considering both target contexts and short-term user preferences. Following this, we utilize contexts aligned with these prototypes to generate a sub-sequence, aimed at predicting CTR and CTCVR scores with target attention. 

Since its inception in 2023, this strategy has been adopted in Meituan Waimai's display recommender system, leading to a 4.6\% surge in CTR and a 4.2\% boost in GMV.

\end{abstract}






\maketitle

\input{intro}

\input{related}

\input{methodology}

\input{exp}

\section{Conclusion}
In this paper, we tackle the challenge of long sequences in the Meituan Waimai recommender system. The main contribution of our work, $\baby$, lies in its integration of context information and the introduction of a fast candidate-agnostic two-stage approach to address this problem. Specifically, $\baby$ introduces a similarity measure based on JS divergence, which enhances interpretability and accuracy. Moreover, it leverages GNN to incorporate temporal information into node representation. Additionally, $\baby$ combines short-term interests to retrieve prototypes that match the target context and utilizes contexts with similar user preferences to obtain a relevant sub-sequence. Our experiments, conducted in both offline and online settings, demonstrate the effectiveness of our proposed model. To the best of our knowledge, we are the first to leverage context information for modeling long sequences in this domain.

\begin{acks} 
This research work was supported by General Program of the National Natural Science Foundation of China (No.62372059). We would like to thank the anonymous reviewers for
their valuable comments.
\end{acks}

% \clearpage
\bibliographystyle{ACM-Reference-Format}
\bibliography{software}

\end{sloppypar}
\end{document}

%% file: intro.tex
\section{INTRODUCTION}

In the intricate ecosystem of Meituan Waimai, China's premier local retail and instant delivery platform, the deployment of an advanced, location-based recommender system is crucial for connecting millions of users with an array of services, from food delivery to pharmaceuticals. The task is made more challenging by the fact that 27\% of users have engaged with the app over 1,000 times in the last year, necessitating a highly efficient process for managing extensive user behavior sequences. Traditional models such as RNNs ~\cite{gru, nasr} falter in maintaining long-term memory across these sequences, while attention-based methods ~\cite{sasrec, bert4rec} are hindered by the significant computational demands of processing Meituan Waimai's long user behavior sequences.

\begin{figure}[!]
  \centering
  \small
  \includegraphics[width=0.5\textwidth]{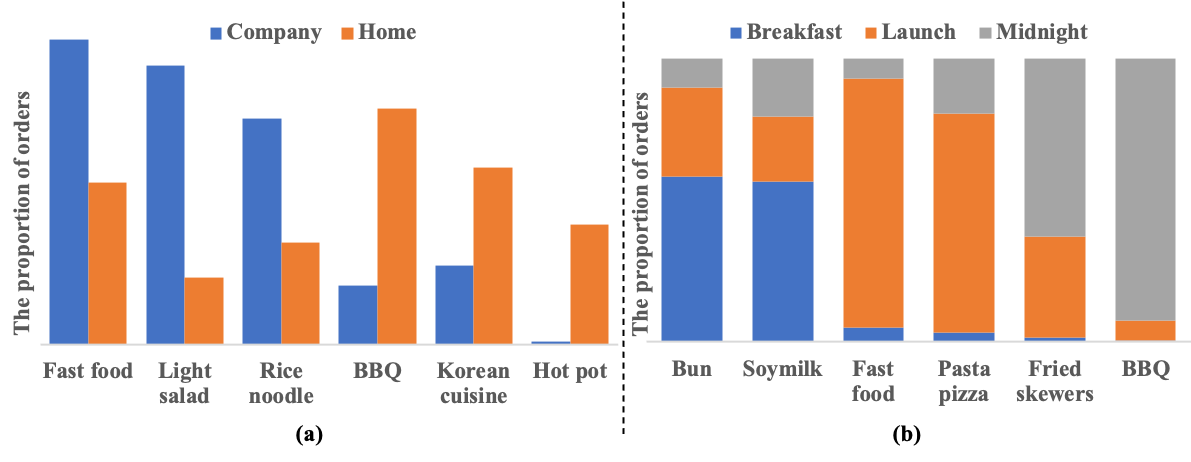}
  \caption{A User's behaviors vary in two characteristics. (a) Geographical location: user interests in different locations (such as companies, and homes) are quite different. (b) Dining time: preference of users at breakfast time (e.g. bun) is different from that at lunchtime (e.g. fast food) and midnight time (e.g. BBQ).}
  \label{fig:intro}
   \vspace*{-0.5\baselineskip}
\end{figure}

Recently, two-stage recommendation models \cite{sim, eta, twin}  have emerged. Based on the target item, the first stage filters out a sub-sequence, and the second stage performs fine-grained modeling on this sub-sequence. However, these models primarily rely on static features such as item attributes for sequence filtering, without adequately considering the dynamic factors that influence user preferences, such as the context during user interaction.
Furthermore, the target attention-based approach integral to these models necessitates traversing the entire sequence for each candidate item, incurring substantial computational overhead. Complicating matters further, these methods rely on the semantic similarity between item embeddings to infer user preferences—a process that can be compromised by the quality of the embeddings, thus affecting the reliability and efficiency of recommendations. In fact, the recommendation process in Meituan Waimai is uniquely influenced by contextual factors like location, time, and weather, diverging from the attribute-centric focus common in e-commerce ~\cite{li2022adver, uriel2022sequential}. This context-driven user preference landscape underscores the inadequacy of traditional methods. Analysis of user interaction data shown in Fig. \ref{fig:intro} reveals that preferences vary considerably across different contexts: for instance, the preference for lighter meals like salads or rice noodles is markedly higher at company, while heartier options such as BBQ and hot pot are favored at home. Mealtime further influences choices, with buns and soymilk popular for breakfast, and fast food or pasta preferred for lunch, transitioning to fried skewers and BBQ for dinner.

To overcome these challenges, in this work, we adhere to the two-stage paradigm, concentrating on the first stage which involves selecting a sub-sequence from the user's behavior sequence. Considering the context-dependency characteristic of Meituan Waimai, we propose to select PoIs that are interacted with under the target context from the user's historical behaviors, which is a method independent of candidate PoIs. However, if we only retain behaviors that completely align with the target context, the retrieved sub-sequence will be excessively short, thereby limiting the effectiveness of the recommendation system.  
For this purpose, we propose a \textbf{Co}ntext-based \textbf{Fa}st \textbf{R}ecommendation \textbf{S}trategy (denoted as \baby) to address long sequence issues in the Meituan Waimai recommender system. Our goal is to identify contexts similar to the target context based on user preferences. We employ Jensen–Shannon (JS) divergence to measure the similarity of PoI attribute distributions across different contexts, enhancing interpretability and accuracy. To reduce complexity, we use a prototype-based approach, where prototypes act as centroids of semantically similar neighbors in the preference representation space. Since prior knowledge of prototypes is not available, inspired by transfer learning ~\cite{transferlearning}, we convert latent representations into probability distributions using an encoder and align them with the actual JS divergence between different contexts obtained from log data using MSE loss. We also build a temporal graph of prototype and context nodes to introduce temporal information. By taking into account the user's short-term preference, we select the prototypes that align with the target context and retrieve the relevant sub-sequence from the user behavior sequence.

To verify the effectiveness of the proposed model, we conduct both offline evaluation experiments on a large user-logged dataset and online A/B testing experiments on the Meituan Waimai app. Experimental results show that our proposed framework outperforms the comparison baselines in both metrics. Our contributions can be summarized as follows:
\begin{itemize}
\item We propose a candidate-agnostic method to model long sequences based on contexts. To the best of our knowledge, this is the first initiative to employ contexts in the modeling of long sequences in this domain.
\item We propose a prototype-based approach to address the challenges of long sequences. Leveraging our proposed probability encoder and graph-based temporal aggregator, we can effectively pinpoint contexts that display preferences akin to the target context.
\item We conduct both offline evaluations on large dataset and online A/B tests on the online platform of Meituan Waimai to demonstrate the effectiveness of our approach.
\end{itemize}

%% file: related.tex
\section{Related Work}
\subsection{Context-aware Recommendation}

Context-aware Recommendation aims to leverage rich context information to improve the recommendation performance ~\cite{norha2018chara}. Firstly, Rendle ~\cite{fm} introduce the factorization machine(FM) to mine the interaction between context features to enhance the recommendation. Inspired by FM, some works ~\cite{fuxing2019interaction, fwfm, wutong2020dual} have been proposed to model the second-order interactions between context features while others ~\cite{xdeepfm, dcnv2} utilize deep neural networks to model high-order interactions between features. Although these methods leverage feature interactions to achieve great performance, their interaction components are empirically predefined. To solve this problem, Cheng et al. ~\cite{weiyu2020adaptive} propose to use logarithmic neural networks (LNN) ~\cite{wesley1996log} to adaptively model the arbitrary-order interactions, while Tian et al. propose EulerNet ~\cite{eulernet} to model the feature interactions in a complex vector space through Euler formula, where the feature interactions in the model are adaptively learned from data.

Recently, several studies also propose to improve the performance of sequential recommendation  by integrating contextual information ~\cite{yang2021light, chang2021non, ren2020sequential, zhang2019feature, context2022ahmed}. However, despite their effectiveness, these methods predominantly rely on the self-attention mechanism, which falls short in addressing the challenges posed by modeling long-term sequences.

\subsection{Long User Behavior Sequence Modeling}
As the sequence lengthens, the model's performance correspondingly enhances. ~\cite{sim}. Existing methods modeling long user behavior sequences can be mainly divided into two categories: RNN-based methods and two-stage methods. In RNN-based methods, some models utilize memory networks to memorize users’ historical behavior sequences ~\cite{hpmn, mimn}, such as HPMN ~\cite{hpmn} and MIMN ~\cite{mimn}. Recently, Wu et al. ~\cite{limarec} propose a linear transformer mechanism based on kernel functions to model long sequences.

Recently, two-stage methods have attracted increasing attention. Pi et al. ~\cite{sim} first propose a two-stage method for long sequence problems, which utilizes a General Search Unit (GSU) module to retrieve items related to the target item from long sequences in hard or soft manners, and then use Exact Search Unit (ESU) module to finely model the retrieved sub-sequence. To speed up the GSU stage, ETA ~\cite{eta} proposes to use locality-sensitive hashing to transform item embeddings into binary vectors, with Hamming distance measuring item similarity. Similarly, with the hashing method, SDIM consolidates and normalizes item embeddings with the same hash as the target, deducing user preference for the target item. However, the relevance between the items retrieved by GSU and  ESU may be relatively low. Therefore, Chang et al.\cite{twin} proposed consistency-preserving GSU (CP-GSU) to enhance the consistency of the two stages. Different from the above retrieval-based GSU method, UBCS ~\cite{ubcs} proposes a clustering-based method to extract representative sub-sequences. In contrast to their studies,our approach involves utilizing context rather than the target item for retrieval during the GSU stage.

%% file: methodology.tex
\section{METHODOLOGY}
In this section, we first introduce the problem formulation of recommendations in Meituan Waimai. We then describe the proposed $\baby$ model in detail.
\subsection{Problem Formulation}
Let $\mathcal{U}$ denote a set of users and $\mathcal{E}$ denote a set of PoIs, where $|\mathcal{U}|$ and $|\mathcal{E}|$ are the numbers of users and PoIs. For each user $u \in \mathcal{U}$, we use $e_{1: n}^u=\left\{e_1^u, e_2^u, \cdots,e_i^u, \cdots, e_n^u\right\}$ to represent the interaction sequence of PoIs, where $e_i^u\in \mathcal{E}$.

Each user interaction in the sequence is accompanied by various contextual features, such as dining time, location, weather, etc. We refer to the combination of these contextual features as a context, like $\langle$ breakfast, office, sunny, $\cdots \rangle$. We denote by \(\mathcal{C}^u\) the set of distinct contexts in which the user has interacted, where \(\mathcal{C}^u = \{\mathcal{C}^u_1, \mathcal{C}^u_2, \cdots, \mathcal{C}^u_{|\mathcal{C}^u|}\}\). Additionally, we use \(\mathcal{A} = \{\mathbf{a}_1, \mathbf{a}_2, ..., \mathbf{a}_{|\mathcal{A}|}\}\) to represent the set of PoI attributes (e.g., categories, prices, etc.), where \(|\mathcal{A}|\) is the number of attribute fields, and each element within it represents a set of possible values that attribute can take. We discretize the continuous attributes into several buckets. To denoise PoIs unrelated to user preferences based on target contexts, while also avoiding issues of inaccurate modeling due to overly short remaining sequences, our goal is to identify contexts with similar user preferences to the target one, which will then be used for filtering in the user behavior sequence.

For simplicity, we describe the technical details of $\baby$ for a single user $u$, and it is straightforward to extend the formulas to a set of users. Hence, we drop subscript $u$ in the notations for concise presentation.

\subsection{Probability Encoder}

Within the Meituan Waimai recommender system, myriad context features such as geographic location, dining time, weather, and holiday presence significantly influence user preferences. Traditional approaches, including cosine similarity of context embeddings, aimed at identifying contexts with analogous user inclinations and selecting sub-sequences pertinent to the target context, encounter challenges including a lack of interpretability and a reliance on the quality of embeddings.
Indeed, the task of accurately capturing user preferences in a specific context using solely latent representations without supervision poses a significant challenge. 
Nonetheless, we observe that user preferences within particular contexts are more accurately represented through probability distributions across PoI attributes, such as category and price. This insight leads us to employ probability distribution consistency as a measure for comparing preferences across different contexts, thereby enhancing both interpretability and accuracy. A prevalent method involves calculating the Kullback-Leibler (KL) divergence. However, to circumvent its inherent asymmetry, we opted for the symmetric alternative, the Jensen-Shannon (JS) divergence. We use $\mathcal{D}(\mathcal{C}_i) = p(a_1, a_2, \cdots, a_{|\mathcal{A}|}| \mathcal{C}_i)$ to denote the distribution of attribute values $\{a_1, a_2, \cdots, a_{|\mathcal{A}|}\}$ under context $\mathcal{C}_i$, which is a $|\mathcal{A}|$-dimension vector recording the proportions of corresponding discrete attribute values that could be obtained from the log data.The equation of divergence between given context $\mathcal{C}_i$ and context $\mathcal{C}_j$ comes as follows: 

\begin{align}
\label{eq:explicit KL}
\mathrm{KL}(\mathcal{C}_{i}, \mathcal{C}_{j}) & = \sum_{{a_1} \in \mathbf{a}_1} \sum_{{a_2} \in \mathbf{a}_2} \cdots \sum_{a_{|\mathcal{A}|} \in \mathbf{a}_{|\mathcal{A}|}} \mathcal{D}(\mathcal{C}_i) \log \frac{\mathcal{D}(\mathcal{C}_i)}{\mathcal{D}(\mathcal{C}_j)} \\
\label{eq:js}
\mathrm{JS}(\mathcal{C}_i, \mathcal{C}_j) &= \frac{1}{2}\left(\mathrm{KL}(\mathcal{C}_i, \mathcal{C}_j) + \mathrm{KL}(\mathcal{C}_j, \mathcal{C}_i)\right)
\end{align}

\begin{figure}[htbp]
\centering
\small
\includegraphics[width=0.45\textwidth]{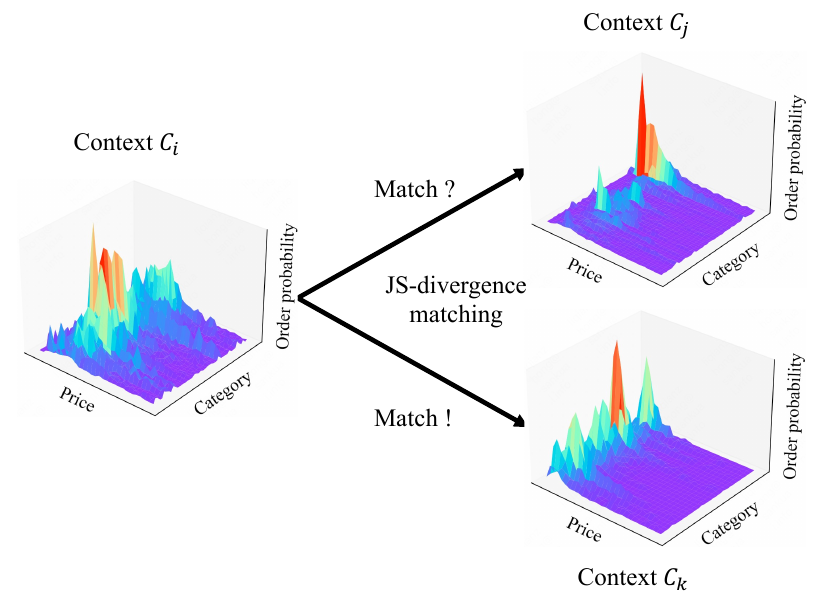}
\caption{An example of the JS divergence calculation based on the PoI attributes between different contexts.}
\label{fig:js}
% \vspace*{-0.2\baselineskip}
\end{figure}

For clarity, the matching process is shown in Fig. ~\ref{fig:js}  by visualizing the distribution of order probability over the price and category dimensions. The order probability is calculated by log data, and it can be viewed as user's interest in PoI in the context.

The described method involves calculating the similarity in preferences between the target context and historical contexts, followed by the selection of a subset of the top similar contexts. However, this strategy is not without its limitations, particularly in its inability to address the challenge of context cold-start. Moreover, the effectiveness of the model is substantially affected by the quantity of contexts selected for each target context, a figure that fluctuates unpredictably across various target contexts.
In fact, it is important to recognize that preferences derived from prolonged behavior within a particular context tend to remain relatively stable.
Therefore, we introduce the concept of preference prototypes, drawing inspiration from prototype learning methodologies referenced in prototype learning literature~\cite{prototype}. We define the collection of these prototypes as \(\mathcal{O} = \{o_1, o_2, \cdots, o_{|\mathcal{O}|}\}\), where \(|\mathcal{O}|\) represents the total number of prototypes. These prototypes serve as the centroids within the preference representation space, grouping semantically akin neighbors. Contexts whose preferences closely match the same prototype are considered analogous. However, as our understanding of the prototypes is limited to their latent representations and we lack explicit prior knowledge about them, directly deriving their probability distributions across PoI attributes is not feasible.
To calculate the JS divergence between prototypes and context preferences, inspired by transfer learning, we design a probability encoder that maps latent representations into probability distributions over PoI attributes. We align the distributions with the ground truth JS divergence between contexts calculated based on log data. Specifically, we use $\hat{\textbf{c}}_i  \in \mathbb{R}^{d}$ to represent the latent representation of the global preference under context $\mathcal{C}_i$, where $d$ represents the number of representation dimensions, and it is shared by all users to reduce the number of parameters. The user's personalized preference is denoted as $\mathbf{c}_i = \mathbf{u} + \hat{\textbf{c}}_i$. To approximate the probability distribution, we employ \(\mathbf{P}(\mathbf{c}_i) = \mathrm{MLP}(\mathbf{c}_i)\), utilizing a Multilayer Perceptron (MLP) model that incorporates a sigmoid activation function. The MLP's output dimensionality is meticulously set to match the aggregate of value counts across all attributes. Based on the estimated probability distributions, we define the estimated JS divergence between representations $\mathbf{c}_i$ and $\mathbf{c}_j$ as follows:

% \begin{small}
\begin{align}
\label{eq:fake_kl}
\widetilde{\mathrm{KL}}(\mathbf{c}_i, \mathbf{c}_j) &= \sum_{k=1}^d c_{i,k} \cdot \log \frac{c_{i,k}}{c_{j,k}} \\
\widetilde{\mathrm{JS}}\left(\mathbf{c}_i, \mathbf{c}_j\right) &= \frac{1}{2} (\widetilde{\mathrm{KL}}({\mathbf{P}(\mathbf{c}}_i), \mathbf{P}({\mathbf{c}_j)) + {\widetilde{\mathrm{KL}}}({\mathbf{P}(\mathbf{c}}_j), \mathbf{P}(\mathbf{c}}_i)))
\end{align}
%\end{small}

where $c_{i,k}$ represent the $k$-th element of $\mathbf{c}_i$. We aim to encourage the estimated JS divergence to approximate the ground truth value define in Eq. ~\ref{eq:explicit KL} obtained from log data. To achieve this, we add the following constraint:

\begin{equation}
\label{eq:mse}
%\small
\mathcal{L}_{MSE} = \frac{1}{{|\mathcal{C}|}^2}\sum_{i=1}^{|\mathcal{C}|} \sum_{j={1}}^{|\mathcal{C}|} {(\mathrm{JS}\left(\mathcal{C}_i, \mathcal{C}_j\right) - \widetilde{\mathrm{JS}}(\mathbf{c}_i, \mathbf{c}_j))}^2
\end{equation}

By minimizing $\mathcal{L}_{MSE}$, we can align the estimated JS divergence and the ground truth. In consequence, we could directly calculate the similarity between latent representations:

\begin{equation}
\label{eq:sim}
    sim(\mathbf{c}_i, \mathbf{c}_j) = 1 - \widetilde{\mathrm{JS}}(\mathbf{c}_i, \mathbf{c}_j)
\end{equation}

It's noteworthy that contexts are dynamically clustered, and surplus prototypes contain virtually no contexts. Consequently, when the number of prototypes surpasses a specific threshold, the model's performance remains relatively stable. In other words, the model's efficacy is not solely contingent on this hyperparameter, a fact we have substantiated in the experiments detailed below.

Besides, we can calculate the JS divergence among prototypes using the probability encoder. To avoid excessive concentration of the prototypes, we introduce an independence loss to constrain prototype representations. Similarly, we use $\hat{\textbf{o}}_i \in \mathbb{R}^d$ to represent the representations shared by users, which follow the same distribution as the representations of context preference, and $\mathbf{o}_i = \hat{\textbf{o}}_i + \mathbf{u}$ to represent personalized representation for prototype $o_i$.To promote the distinctiveness of different prototypes, we encourage their separation through JS divergence:

%\begin{small}
\begin{align}
\label{eq:context emb}
\mathcal{L}_{IND} &= - \frac{1}{{|\mathcal{O}|}^2}\sum_{i=1}^{|\mathcal{O}|} \sum_{j={1}}^{|\mathcal{O}|} {\widetilde{\mathrm{JS}}\left(\mathbf{o}_i, \mathbf{o}_j\right)}
\end{align}
%\end{small}

By minimizing $\mathcal{L}_{IND}$, we can make the prototypes more evenly distributed in the latent space.

\subsection{Graph-based Temporal Aggregator}
While the probability encoder effectively clusters contexts, it overlooks the temporal dynamics embedded within sequences. This oversight becomes particularly critical in handling long sequences where traditional RNN-based models may grapple with memory loss over time, and the computational demand of attention-based models becomes prohibitively intricate.
Recently, graph neural networks have been applied to sequential recommendation, constructing connected graphs of items to handle long user behavior sequence ~\cite{sugre}, modeling long-term item dependencies. A common method for graph construction utilizes the co-occurrence relationships between items to establish connections. However, due to the sparsity of items in user behavior sequence, the graphs often degenerate into linear structures, limiting the effectiveness of GNNs. Consequently, GNN applications are mainly found in areas with frequent item repetition, like session-based recommendations~\cite{SRGNN}. Nevertheless,  within context sequences, the restricted diversity of contexts and the extensive sequence length lead to repeated contexts, forming an optimal scenario for GNN modeling. Accordingly, a temporal graph is constructed based on the sequential order of contexts, wherein nodes denote contexts, and edges signify the co-occurrences among adjacent contexts. This framework proficiently encapsulates the inherent sequential dynamics within user behavior, offering a nuanced approach to understanding and modeling user interactions.

\begin{figure}[htbp]
\centering
\includegraphics[width=0.45\textwidth]{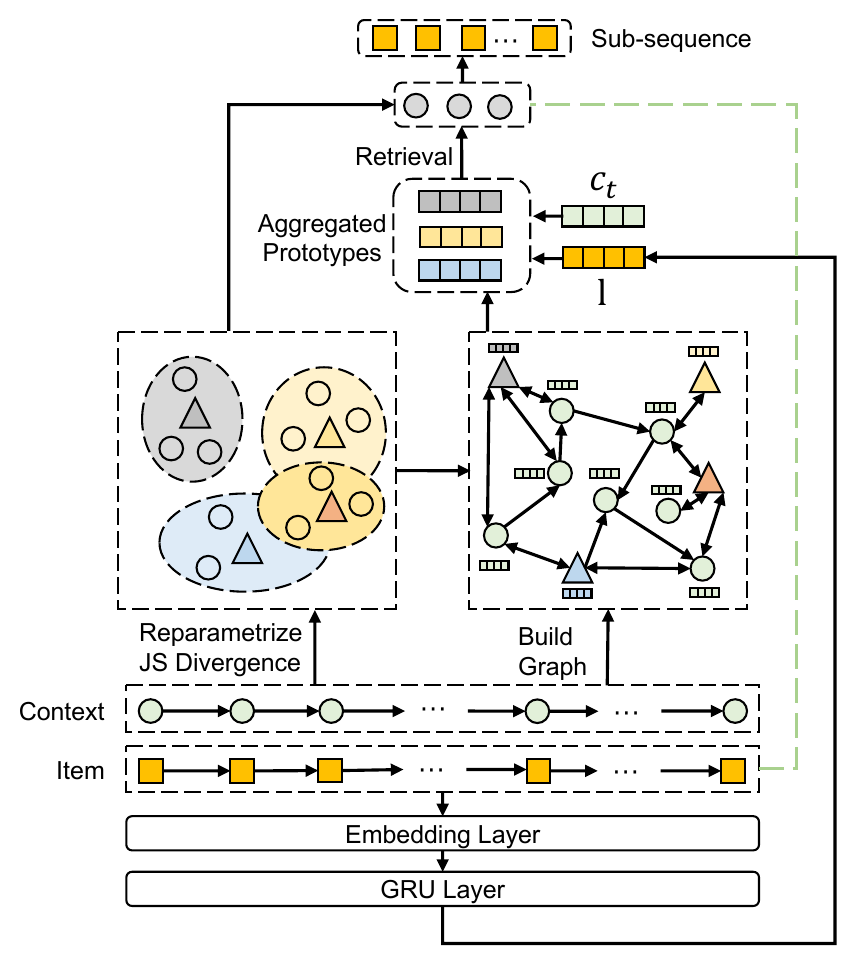}
\caption{The overall architecture of Context-based User fast Recommendation Strategy($\baby$ for short) for long sequences in recommender system of Meituan Waimai.}
% \vspace{-0.1cm}

\label{fig:model}
\end{figure}

To incorporate temporal dimensions into prototypes, we embed prototype nodes within the graph. However, fully connecting prototype and context nodes risks creating an unwieldy graph size. To mitigate this, we suggest a filtering method that retains only edges with similarity scores above a certain threshold. However, the conventional hard-coding filtering mechanism is not differentiable, obstructing effective backpropagation training. To bypass this, we adopt the Gumbel Softmax trick from prior work \cite{gumbel}, facilitating differentiable learning over discrete outputs. The equation for each prototype $i$ and context $j$ is as follows:

%\begin{small}
\begin{align}
\label{eq:gumbel}
\beta &= sim(\mathbf{o}_i, \mathbf{c}_j) \\
\mathcal{P}(\beta ) & =\frac{\exp ((\log (\beta + g_b) / {\tau_p})}{\sum_{b=0}^1 \exp (\log (\beta^b(1-\beta)^{1-b})+g_b) / {\tau_p})}
\end{align}
%\end{small}
where $g_b$ represents a noise sampled from a Gumbel distribution, and the temperature parameter $\tau_p$ controls its sharpness. Additionally, the similarity between each prototype-context pair is pre-calculated and saved in the storage, eliminating the need for real-time computation online.
For the pair where $\mathcal{P}\left(\beta\right)=1$, we add a bidirectional edge between the nodes. Next, we utilize Graph Attention Network \cite{gat}, which could deal with directed graphs and unseen nodes, to aggregate the graph and embed temporal information into the node representation:

\begin{small}
    \begin{align}
    \label{eq:gnn}
    \mathbf{H}^0 &= [\mathbf{o}_1 ; \cdots ; \mathbf{o}_{|\mathcal{O}|} ; \mathbf{c}_1 ; \cdots ;\mathbf{c}_{|\mathcal{C}|} ] \\
    \alpha_{i j}^l &=\frac{\exp (\operatorname{LeakyReLU}(sim(\mathbf{H}_i^l \mathbf{W}^l,  \mathbf{H}_j^l \mathbf{W}^l))}{\sum_{k \in \mathcal{N}_i} \exp (\operatorname{LeakyReLU}( sim(\mathbf{H}_i^l \mathbf{W}^l, \mathbf{H}_k^l \mathbf{W}^l)))} \\
    \mathbf{H}_i^{l+1} &=\sigma(\sum_{j \in \mathcal{N}_i} \alpha_{i j} \mathbf{H}_j^l \mathbf{W} ).
    \end{align}
\end{small}

where $\textbf{H}^{l}_i \in \mathbb{R}^d$ and $\mathbf{W}^{l} \in \mathbb{R}^{d\times d}$ represent the $i$-th node representation and  the parameters in the $l$-th layer, respectively. $\sigma$ represents the ReLU activation function. $\mathcal{N}_i$ denotes the predecessor nodes of node $i$.After $L$ times of aggregation, we can obtain the $i$-th aggregated prototype representation $\widetilde{\mathbf{o}_i} = \mathbf{H}^L_{i}$ and $j$-th context preference representation $\widetilde{\mathbf{c}}_{j} = \mathbf{H}^L_{|\mathcal{O}|+j}$ integrated with temporal information.

\subsection{Learning and Discussion}
% \subsubsection{Learning}

Once having the aggregated node representations, we identify the prototypes that fit with the target context by focusing on the user's most recent $r$ activities $\mathbf{e}_{(n-r):n} = \{\mathbf{e}_{(n-r)}, \mathbf{e}_{(n-r+1)}, \cdots, \mathbf{e}_n\}$, where $r\ll n$. We encode the user's short-term behavior using GRU model \cite{gru}:

\begin{equation}
%\small
\label{eq:gru}
    \mathbf{l} = \mathrm{GRU}(\mathbf{e}_{(n-r)}, \mathbf{e}_{(n-r+1)}, \cdots, \mathbf{e}_{n})
\end{equation}

where $\mathbf{l}$ represents the user's short-term preference. Next, we assume that the target context of the user is $\mathcal{C}_t$.Then, we calculate the relation between prototypes and the target context:

\begin{equation}
    \label{eq:flag}
    z_i^t = \mathcal{P}\left(sim(\widetilde{\mathbf{o}_i}, \widetilde{\mathbf{c}_{t}}){(\widetilde{\mathbf{o}_i} \cdot \mathbf{l})}\right)
\end{equation}

where the dot represents inner product operation. As there is no concept of probability distribution on attributes for PoIs, we compute the similarity between prototypes and candidate PoI using the inner product. To ensure effective training, we utilize binary cross entropy loss as Eq. ~\ref{rec} to provide supervision for the entire process.

\begin{align}
    %\small
    \label{rec}
    s_{i,v}^t &=  z_i^t \cdot (\widetilde{\mathbf{o}_i} \cdot \mathbf{v}) \\
    \mathcal{L}_{REC} &= \frac{1}{|V|}\sum_{\mathbf{v} \in V}\sum_{i=1}^{|\widetilde{\mathcal{O}}|} [-y \cdot \log{s_{i,v}^t} - (1-y) \cdot \log(1 - {s_{i,v}^t})]
\end{align}
where $V$ denotes the training set, and $y=1$ for positive samples while $y=0$ otherwise. The final loss is as follows:

\begin{equation}
\label{eq:final_loss}
    \mathcal{L} = \mathcal{L}_{REC} + \gamma \cdot \mathcal{L}_{MSE} + \lambda \cdot \mathcal{L}_{IND}
\end{equation}

where $\gamma$ and $\lambda$ control the weights of the MSE loss and the independence loss, respectively. The overall learning algorithm for the proposed method is given in Alg.~\ref{alg:learning_state}. 

\begin{small}
\begin{algorithm}[htb]
\caption{Learning algorithm for $\baby$}
\label{alg:learning_state}
\renewcommand{\algorithmicrequire}{\textbf{Input:}}
\renewcommand{\algorithmicensure}{\textbf{Output:}}
\begin{algorithmic}[1]
\REQUIRE  Training epoch ${T}_{train}$
    \STATE Initialize model parameters ${\Theta}$ $\leftarrow$ random values;
    \FOR {$T_{train}$ epochs}
        
        \STATE Construct a temporal graph of contexts based on the user's historical behavior;
        \STATE Calculate the similarity between prototypes and contexts according to Eq. \ref{eq:sim};
        \STATE Integrate prototype nodes into the graph, establishing connections between prototypes and contexts according to Eq. \ref{eq:gumbel};
        \STATE Perform message passing between nodes based on Eq. \ref{eq:gnn}; 
        \STATE Derive the user's short-term preferences through Eq.\ref{eq:gru};
        \STATE Identify the prototypes that match the target context using Eq. ~\ref{eq:flag};
        \STATE Optimize the model parameters by minimizing the loss function Eq. ~\ref{eq:final_loss};
    \ENDFOR
    \RETURN  All optimal parameters in $\Theta$.

\end{algorithmic}
\end{algorithm}
\end{small}

In the next phase, for each prototype \(o_i\) in the set \(\mathcal{O}\) where \(z_i^t = 1\), we collect the contexts that are encompassed by it. Within these contexts, we identify and select a sub-sequence of PoIs with which the user has engaged. This selected sub-sequence is then processed using target attention~\cite{din}, to intricately model and capture user interest. The model is trained under the cross-entropy loss function.

For the online prediction of a user's subsequent visit within the target context \(\mathcal{C}_t\), we implement diverse strategies. For previously encountered context, we use the pre-calculated similarity between prototype-context pairs. For cold-start context, we employ the representation \(\mathbf{c}_t\) to calculate its similarity with the prototypes. Based on this representation, we emulate the learning process mentioned in this section to select sub-sequence and model user interest.

% \paratitle{Complexity Analysis}
The proposed $\baby$ method selects a sub-sequence based on contexts with similar preferences to the target context independent of candidate items. In this section, we analyze the time complexity of inference time in the GSU stage. In the offline phase, it calculates the similarity between prototypes and context preferences with a time complexity of O(|$\mathcal{C}| \cdot |\mathcal{O}$|), where $|\mathcal{C}|$ is much smaller than the sequence length $n$, and $|\mathcal{O}|$ is a constant. The time complexity of graph aggregation is $O(n \cdot L)$. In the online phase, we need $O(r + |\mathcal{O}|)$ to model the short-term preference to obtain the selected sub-sequence and use it to retrieve relevant prototypes.  It is evident that our approach has a significantly better time complexity compared to other two-stage models, which have a time complexity of $O(B \cdot n)$, where $B$ represents the number of candidates.

%% file: exp.tex
\section{Experiment}
\subsection{Experiemental Setup}

\paratitle{Dataset.} For offline experiments, we collect training data by log data of Meituan Waimai from April 1st to April 30th, 2023, and the validation data is constructed by the samples on May 9th, 2023. Especially, sample data contains user profile/statistical features, statistical features of PoIs, raw user-behavior features, labels (whether click or not), and so on. The raw user-behavior features include all PoIs that users have interacted with over the past three years. In addition, we use context features including geographical location, mealtime, weather, and the presence of holidays. For PoI attributes, we consider category, price, quality, and delivery time. The statistics of our dataset are shown in Tab. ~\ref{t:data}, where Recent Avg. refers to the average click records number of users in the training set, while Historical Avg. refers to the average interaction records number of users over the past three years, which are used to model user interests.

\begin{table}[htb]
\caption{Statistics of the dataset}
\label{t:data}
\begin{tabular}{cc}
\hline
Field                                                                         & Size         \\ \hline
\#Users                                                                       & 0.24 billion \\
\#PoIs                                                                       & 4.37 million \\
\#Records                                                                     & 36 billion   \\
\#Recent Avg.    & 150          \\
\#Historical Avg. & 4,423          \\ \hline
\end{tabular}
\end{table}

\paratitle{Baseline.} To evaluate the effectiveness of our approach, we compare our model against two types of baselines, including three traditional models, and five enhanced models for long sequences. Specifically, we consider three traditional models:
\begin{itemize}
\item {\textbf{Avg-Pooling}~\cite{youtube}} is a basic deep learning model for CTR prediction. It performs average pooling to integrate behavior embeddings.
\item {\textbf{DIN}\cite{din}} is an attention-based method that only utilizes short-term user behavior sequences to model user interests.
\item {\textbf{DIEN}\cite{dien}} integrates GRU with a candidate-based attention mechanism to model user interests.
\end{itemize}
For enhanced models for long sequences, we consider the following five baselines:
\begin{itemize}
\item {\textbf{MIMN\cite{mimn}}}: utilizes a memory network to capture long-term user interest.
\item{\textbf{SIM}$_{hard}$\cite{sim}} gets relevant interacted PoIs by category id.
\item {\textbf{SIM}$_{soft}$\cite{sim}} searches top-k interacted PoIs by the similarity of PoI embeddings. For our experiments, the well-trained PoI embeddings are obtained from our Recall System.
\item {\textbf{SDIM} \cite{sdim}} aggregates and normalizes the embeddings of items with the same hash as the target item in the user behavior sequence to derive the user preference related to the target item.
\item {\textbf{TWIN}}~\cite{twin} is an improved version of SIM, which enhances the consistency of embeddings in GSU and ESU stages.
\end{itemize}

\paratitle{Parameter Setting.} For a fair comparison, we adopt the following setting for all methods: the number of training epochs is set to 500 and the batch size is set to 128. All embedding parameters are randomly initialized in the range of (0, 1), and the learning rate is tuned in the range of \{1e-4, 5e-4, 1e-3\} and the embedding size is tuned in the range of \{8, 16, 32\}. For other parameters, we follow the settings in the original paper or source code.

%TODO: batch size?

For $\baby$, we implement it based on PyTorch with Adam optimizer. For Eq. ~\ref{eq:gumbel}, the $\tau_p$ is set to tuned in \{1e-3, 1e-2, 0.1, 1, 2\}. The $\gamma$ and $\lambda$ in Eq.~\ref{eq:final_loss} are set to 5e-2 and 1e-3, respectively. And the number of prototypes is tuned in \{1, 10, 20, 30, 40, 60, 80\}, which is studied in the subsequent section. 

\subsection{Performance Comparison}

In this section, we compare the performance of our model with the baselines. The overall performance of our proposed $\baby$ and the baselines are reported in Tab. ~\ref{t:main}. We have the following observations:

For traditional models, we observe that Avg-Pooling has the poorest performance. By employing target attention to assign varying weights to PoIs in the behavior sequence, we discover that DIN and DIEN outperform models without attention, demonstrating the significance of diminishing the impact of irrelevant PoIs. Furthermore, by incorporating temporal information, DIEN outperforms DIN, which validates the importance of integrating sequential information in $\baby$. 

In terms of enhanced recommenders for long sequences, their performance surpasses that of DIN and DIEN, which are short-term models. This indicates the critical role of long sequences in more precisely capturing user interests. We also observed that MIMN has the least effective performance among all enhanced models. This could be due to its approach of updating memory by assigning weights to PoIs in the sequence via a soft attention manner, where potential noise accumulation can significantly hinder the sequence learning process. Consequently, its performance lags behind two-stage methods that employ a hard-coding approach. A similar observation could be explored in ~\cite{rap, clea}. Additionally, despite a reduction in time complexity, SDIM outperforms SIM$_{hard}$ and exhibits comparable performance to SIM$_{soft}$, validating the efficacy of the hashing method proposed by SDIM. Furthermore, TWIN excels over SDIM as it not only enhances the consistency of the two-stage representation but also employs target attention instead of normalizing the PoI embeddings for sequential modeling, thereby amplifying the correlation between user interest and the target PoI.

Finally, $\baby$ outperforms all baselines and outperforms TWIN by 0.77\% on CTR AUC and 0.63\%  on CTCVR AUC, respectively, which proves the effectiveness of the proposed $\baby$. By introducing contextual information, $\baby$, with the proposed probability encoder and graph-based temporal aggregator, $\baby$ can effectively and efficiently identify contexts with similar preferences, according to which it selects PoIs relevant to the target context.

\begin{table}[!h] 
\caption{Performance comparison between our \baby and baselines. The best performance of each field is highlighted in boldface. The improvement of \baby against the best baseline is consistently significant at 0.05 level.}

\centering
\label{t:main}
\begin{tabular}{c|ccc}
\toprule
\multicolumn{1}{l|}{}  & Model           & \multicolumn{1}{l}{CTR AUC} & \multicolumn{1}{l}{CTCVR AUC} \\ \midrule
\multirow{3}{*}{Traditional Models}                                                           
& Avg-Pooling & 0.6993                      & 0.7204                        \\
& DIN             & 0.7052                      & 0.7263                        \\
& DIEN            & 0.7069                      & 0.7281                        \\ \hline
\multirow{6}{*}{\begin{tabular}[c]{@{}l@{}}Enhanced Models \\ for Long Sequences\end{tabular}} 
& MIMN            & 0.7083                      & 0.7287                        \\
& SIM$_{hard}$             & 0.7097                      & 0.7300                        \\
& SIM$_{soft}$             & 0.7116                      & 0.7321                        \\
& SDIM            & 0.7123                      & 0.7325                        \\
& TWIN            & 0.7133                      & 0.7336                        \\
& $\baby$          & \textbf{0.7188}                      & \textbf{0.7382}                        \\ \bottomrule
\end{tabular}
\end{table}

\subsection{Ablation Study}
Our proposed $\baby$ method measures the similarity between the prototypes and the context preferences based on JS divergence and constrains the latent representations through several losses while introducing temporal information into the representation through GNN.To verify the effectiveness of such a design, we consider the following variants of our model for comparison through offline experiments:

\begin{itemize}
    \item {$\baby_{IP}$}: This variant represents using the inner product instead of JS divergence as the similarity measure. 
    \item {$\baby_{\neg{MSE}}$}: This variant represents removing $\mathcal{L}_{MSE}$ for comparison with $\baby$ and $\baby_{IP}$.
    \item {$\baby_{\neg{IND}}$}: This variant represents removing the independence constraint of prototypes.
    \item {$\baby_{\neg{GTA}}$}: This variant represents removing the graph-based temporal aggregator and directly using the original representations for subsequent operations instead of the aggregated ones.
\end{itemize}

\begin{table}[!h]
\caption{{Performance comparison of $\baby$ and different variants}
}
\centering
\label{t:ablation}
\begin{tabular}{cccccccc}
\toprule
Model & CTR AUC & CTCVR AUC\\
\hline
$\baby_{IP}$ &0.7146  &0.7341 \\
$\baby_{\neg{MSE}}$ &0.7128 &0.7326 \\
$\baby_{\neg{IND}}$ &0.7149 &0.7343 \\
$\baby_{\neg{GTA}}$ &0.7141 & 0.7339 \\
\baby & \textbf{0.7188} & \textbf{0.7382}\\
\hline
\end{tabular}
\end{table}

The performance comparison among $\baby$ and its variants is shown in Tab. ~\ref{t:ablation}. From the table, we have the following observations. Firstly, regarding the similarity measure, we found that using the inner product instead of the similarity defined in Eq. \ref{eq:sim} leads to a decrease in performance. This proves that calculating JS divergence based on the attribute distributions of the context indeed brings a more accurate similarity measure. In addition, if we only remove the alignment constraint in Eq. \ref{eq:mse}, this will lead to poor performance, which indicates that it is difficult for the model to find contexts with similar preferences by calculating JS divergence based on random and unaligned representations. Secondly, we found that removing $\mathcal{L}_{IND}$ leads to a degradation in performance as well. The underlying reason might be that without the independence constraint, the distribution of each prototype becomes concentrated, making it challenging for the model to distinguish between diverse user preferences. Thirdly, we observe a performance decline when the graph-based temporal aggregator is removed. This confirms the importance of integrating transit relationships between contexts into the representations, as its absence leads to a loss of temporal information.

\subsection{Analysis on Effect of Prototype Number}
\begin{figure}[!]
  \centering
  \includegraphics[width=0.46\textwidth]{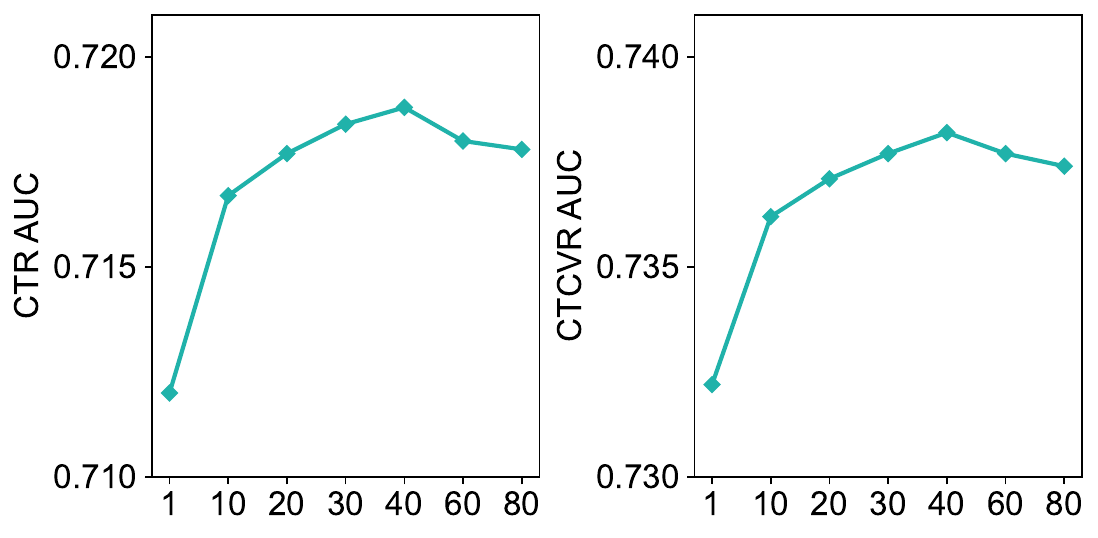}
  \caption{Performance variation of different prototype numbers}
  \label{fig:prototype}
\end{figure}

Recall that in our proposed CoFARS, the number of prototypes is a hyperparameter, in this senction, we evaluate the effectiveness of prototypes in $\baby$ by altering the number of prototypes |$\mathcal{O}$|. The outcomes are depicted in Fig.\ref{fig:prototype}. When $|\mathcal{O}|=1$, we only retain the PoIs that have been interacted with in the target context previously. Remarkably, $\baby$ consistently surpasses the baseline for various |$\mathcal{O}$| values, except when $|\mathcal{O}|=1$. Optimal performance is attained when |$\mathcal{O}$| is approximately 40, suggesting that contexts with similar user preferences are appropriately assigned to prototypes. On the contrary, setting |$\mathcal{O}$| to 1 results in a substantial performance drop. This can be attributed to the filtered sequences are too short and monotonous  to capture users' interest diversity and its evolution.

\subsection{Visualization Analysis}
\begin{figure}[!t]
\centering
\includegraphics[scale=0.5]{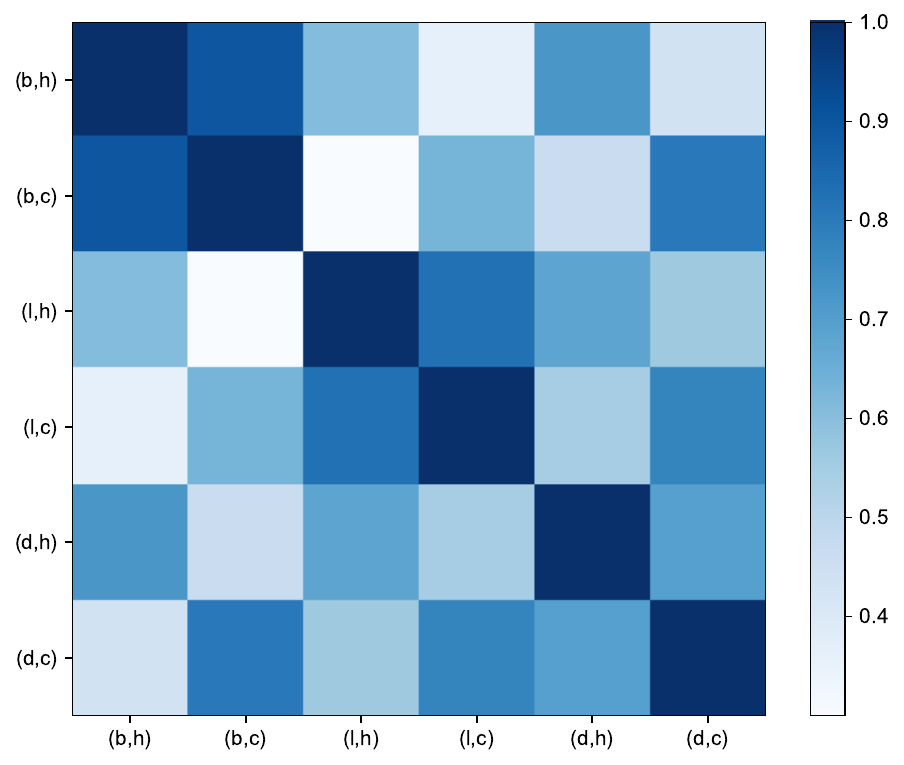}
\caption{Similarity between preferences under different contexts. The x-axis and y-axis both denote the contexts, $b$ stands for breakfast, $l$ stands for lunch, $d$ stands for dinner, $h$ represents home, and $c$ represents company.}
% \vspace{-0.15cm}
\label{fig:case_study}
\end{figure}

In this section, to delve deeper into whether the learned representations of our model can genuinely discern contexts with similar user preferences, we have undertaken a visualization of the similarity among contexts. Specifically, we have selected a user as an example, whose mealtimes encompass breakfast, lunch, and dinner, and whose habitual locations include his home (a residential area) and his company (a business district). By utilizing the Cartesian product, we are able to generate six distinct contexts. Leveraging the well-trained embedding, we employ the Eq. ~\ref{eq:js} to compute the similarity of user preferences across these contexts and perform normalization. The visualization result of all representations is shown in Fig. ~\ref{fig:case_study}.

In the figure, we've uncovered several insights. Foremost, it is clear that within the same contextual features, users demonstrate similar preferences. For example, when mealtime is constant while the location changes, the context similarity is higher than in completely unrelated contexts. Specifically, $sim(\langle b,h \rangle, \langle b,o \rangle)$ exceeds  $sim(\langle b,h \rangle, \langle l,o \rangle)$ and $sim(\langle b,h \rangle, \langle d,o \rangle)$. This pattern is consistent across all mealtimes. With a static location, similar trends are observed, which is likely attributable to consistent preferences for the same mealtime or location. For instance, 
users may prefer heavier lunches and lighter meals for breakfast and dinner, meanwhile choose pricier PoIs at home and cheaper options at the company. 
Secondly, breakfast orders at home and the company are more similar than that of lunch and dinner, possibly due to the limited breakfast variety compared to lunch and dinner. Lastly, the minimum similarity across all contexts exceeds 0.3, indicating basic preferences like favoring lower-priced PoIs.

\subsection{Online A/B Testing}
$\baby$ is conducted in the recommender system of Meituan Waimai for online A/B testing from 2023-05 to 2023-06. During almost a month of testing, compared with the baseline, the last version of our online-serving model, $\baby$ obtained 4.6\% CTR and 4.2\% GMV promotion. The significant improvement demonstrates the effectiveness of our proposed approach. $\baby$ has been deployed online and serving the main traffic in our real system now.

%% file: main.bbl

\begin{thebibliography}{40}

%%% the Web address.

\ifx \showCODEN    \undefined \def \showCODEN     #1{\unskip}     \fi
\ifx \showDOI      \undefined \def \showDOI       #1{#1}\fi
\ifx \showISBNx    \undefined \def \showISBNx     #1{\unskip}     \fi
\ifx \showISBNxiii \undefined \def \showISBNxiii  #1{\unskip}     \fi
\ifx \showISSN     \undefined \def \showISSN      #1{\unskip}     \fi
\ifx \showLCCN     \undefined \def \showLCCN      #1{\unskip}     \fi
\ifx \shownote     \undefined \def \shownote      #1{#1}          \fi
\ifx \showarticletitle \undefined \def \showarticletitle #1{#1}   \fi
\ifx \showURL      \undefined \def \showURL       {\relax}        \fi
% invisible to TeX
\providecommand\bibfield[2]{#2}
\providecommand\bibinfo[2]{#2}
\providecommand\natexlab[1]{#1}
\providecommand\showeprint[2][]{arXiv:#2}

\bibitem[Cao et~al\mbox{.}(2022)]%
        {sdim}
\bibfield{author}{\bibinfo{person}{Yue Cao}, \bibinfo{person}{Xiaojiang Zhou},
  \bibinfo{person}{Jiaqi Feng}, \bibinfo{person}{Peihao Huang},
  \bibinfo{person}{Yao Xiao}, \bibinfo{person}{Dayao Chen}, {and}
  \bibinfo{person}{Sheng Chen}.} \bibinfo{year}{2022}\natexlab{}.
\newblock \showarticletitle{Sampling Is All You Need on Modeling Long-Term User
  Behaviors for {CTR} Prediction}. In \bibinfo{booktitle}{\emph{{CIKM}}}.
  \bibinfo{publisher}{{ACM}}, \bibinfo{pages}{2974--2983}.
\newblock


\bibitem[Chang et~al\mbox{.}(2021)]%
        {sugre}
\bibfield{author}{\bibinfo{person}{Jianxin Chang}, \bibinfo{person}{Chen Gao},
  \bibinfo{person}{Yu Zheng}, \bibinfo{person}{Yiqun Hui},
  \bibinfo{person}{Yanan Niu}, \bibinfo{person}{Yang Song},
  \bibinfo{person}{Depeng Jin}, {and} \bibinfo{person}{Yong Li}.}
  \bibinfo{year}{2021}\natexlab{}.
\newblock \showarticletitle{Sequential Recommendation with Graph Neural
  Networks}. In \bibinfo{booktitle}{\emph{{SIGIR}}}.
  \bibinfo{publisher}{{ACM}}, \bibinfo{pages}{378--387}.
\newblock


\bibitem[Chang et~al\mbox{.}(2023)]%
        {twin}
\bibfield{author}{\bibinfo{person}{Jianxin Chang}, \bibinfo{person}{Chenbin
  Zhang}, \bibinfo{person}{Zhiyi Fu}, \bibinfo{person}{Xiaoxue Zang},
  \bibinfo{person}{Lin Guan}, \bibinfo{person}{Jing Lu}, \bibinfo{person}{Yiqun
  Hui}, \bibinfo{person}{Dewei Leng}, \bibinfo{person}{Yanan Niu},
  \bibinfo{person}{Yang Song}, {and} \bibinfo{person}{Kun Gai}.}
  \bibinfo{year}{2023}\natexlab{}.
\newblock \showarticletitle{{TWIN:} TWo-stage Interest Network for Lifelong
  User Behavior Modeling in {CTR} Prediction at Kuaishou}. In
  \bibinfo{booktitle}{\emph{{KDD}}}. \bibinfo{publisher}{{ACM}},
  \bibinfo{pages}{3785--3794}.
\newblock


\bibitem[Chen et~al\mbox{.}(2021)]%
        {eta}
\bibfield{author}{\bibinfo{person}{Qiwei Chen}, \bibinfo{person}{Changhua Pei},
  \bibinfo{person}{Shanshan Lv}, \bibinfo{person}{Chao Li},
  \bibinfo{person}{Junfeng Ge}, {and} \bibinfo{person}{Wenwu Ou}.}
  \bibinfo{year}{2021}\natexlab{}.
\newblock \showarticletitle{End-to-End User Behavior Retrieval in Click-Through
  RatePrediction Model}.
\newblock   \bibinfo{volume}{abs/2108.04468} (\bibinfo{year}{2021}).
\newblock


\bibitem[Cheng et~al\mbox{.}(2020)]%
        {weiyu2020adaptive}
\bibfield{author}{\bibinfo{person}{Weiyu Cheng}, \bibinfo{person}{Yanyan Shen},
  {and} \bibinfo{person}{Linpeng Huang}.} \bibinfo{year}{2020}\natexlab{}.
\newblock \showarticletitle{Adaptive Factorization Network: Learning
  Adaptive-Order Feature Interactions}. In \bibinfo{booktitle}{\emph{{AAAI}}}.
  \bibinfo{publisher}{{AAAI} Press}, \bibinfo{pages}{3609--3616}.
\newblock


\bibitem[Covington et~al\mbox{.}(2016)]%
        {youtube}
\bibfield{author}{\bibinfo{person}{Paul Covington}, \bibinfo{person}{Jay
  Adams}, {and} \bibinfo{person}{Emre Sargin}.}
  \bibinfo{year}{2016}\natexlab{}.
\newblock \showarticletitle{Deep Neural Networks for YouTube Recommendations}.
  In \bibinfo{booktitle}{\emph{RecSys}}. \bibinfo{publisher}{{ACM}},
  \bibinfo{pages}{191--198}.
\newblock


\bibitem[Hidasi et~al\mbox{.}(2016)]%
        {gru}
\bibfield{author}{\bibinfo{person}{Bal{\'{a}}zs Hidasi},
  \bibinfo{person}{Alexandros Karatzoglou}, \bibinfo{person}{Linas Baltrunas},
  {and} \bibinfo{person}{Domonkos Tikk}.} \bibinfo{year}{2016}\natexlab{}.
\newblock \showarticletitle{Session-based Recommendations with Recurrent Neural
  Networks}. In \bibinfo{booktitle}{\emph{{ICLR}}}.
\newblock


\bibitem[Hines(1996)]%
        {wesley1996log}
\bibfield{author}{\bibinfo{person}{J.~Wesley Hines}.}
  \bibinfo{year}{1996}\natexlab{}.
\newblock \showarticletitle{A logarithmic neural network architecture for
  unbounded non-linear function approximation}. In
  \bibinfo{booktitle}{\emph{{ICNN}}}. \bibinfo{publisher}{{IEEE}},
  \bibinfo{pages}{1245--1250}.
\newblock


\bibitem[Hong et~al\mbox{.}(2019)]%
        {fuxing2019interaction}
\bibfield{author}{\bibinfo{person}{Fuxing Hong}, \bibinfo{person}{Dongbo
  Huang}, {and} \bibinfo{person}{Ge Chen}.} \bibinfo{year}{2019}\natexlab{}.
\newblock \showarticletitle{Interaction-Aware Factorization Machines for
  Recommender Systems}. In \bibinfo{booktitle}{\emph{{AAAI}}}.
  \bibinfo{publisher}{{AAAI} Press}, \bibinfo{pages}{3804--3811}.
\newblock


\bibitem[Jang et~al\mbox{.}(2017)]%
        {gumbel}
\bibfield{author}{\bibinfo{person}{Eric Jang}, \bibinfo{person}{Shixiang Gu},
  {and} \bibinfo{person}{Ben Poole}.} \bibinfo{year}{2017}\natexlab{}.
\newblock \showarticletitle{Categorical Reparameterization with
  Gumbel-Softmax}. In \bibinfo{booktitle}{\emph{{ICLR} (Poster)}}.
  \bibinfo{publisher}{OpenReview.net}.
\newblock


\bibitem[Juan et~al\mbox{.}(2016)]%
        {fwfm}
\bibfield{author}{\bibinfo{person}{Yu{-}Chin Juan}, \bibinfo{person}{Yong
  Zhuang}, \bibinfo{person}{Wei{-}Sheng Chin}, {and}
  \bibinfo{person}{Chih{-}Jen Lin}.} \bibinfo{year}{2016}\natexlab{}.
\newblock \showarticletitle{Field-aware Factorization Machines for {CTR}
  Prediction}. In \bibinfo{booktitle}{\emph{RecSys}}.
  \bibinfo{publisher}{{ACM}}, \bibinfo{pages}{43--50}.
\newblock


\bibitem[Kang and McAuley(2018)]%
        {sasrec}
\bibfield{author}{\bibinfo{person}{Wang{-}Cheng Kang} {and}
  \bibinfo{person}{Julian~J. McAuley}.} \bibinfo{year}{2018}\natexlab{}.
\newblock \showarticletitle{Self-Attentive Sequential Recommendation}. In
  \bibinfo{booktitle}{\emph{{ICDM}}}. \bibinfo{pages}{197--206}.
\newblock


\bibitem[Li et~al\mbox{.}(2017)]%
        {nasr}
\bibfield{author}{\bibinfo{person}{Jing Li}, \bibinfo{person}{Pengjie Ren},
  \bibinfo{person}{Zhumin Chen}, \bibinfo{person}{Zhaochun Ren},
  \bibinfo{person}{Tao Lian}, {and} \bibinfo{person}{Jun Ma}.}
  \bibinfo{year}{2017}\natexlab{}.
\newblock \showarticletitle{Neural Attentive Session-based Recommendation}. In
  \bibinfo{booktitle}{\emph{{CIKM}}}. \bibinfo{publisher}{{ACM}},
  \bibinfo{pages}{1419--1428}.
\newblock


\bibitem[Li et~al\mbox{.}(2021b)]%
        {prototype}
\bibfield{author}{\bibinfo{person}{Junnan Li}, \bibinfo{person}{Pan Zhou},
  \bibinfo{person}{Caiming Xiong}, {and} \bibinfo{person}{Steven C.~H. Hoi}.}
  \bibinfo{year}{2021}\natexlab{b}.
\newblock \showarticletitle{Prototypical Contrastive Learning of Unsupervised
  Representations}. In \bibinfo{booktitle}{\emph{{ICLR}}}.
  \bibinfo{publisher}{OpenReview.net}.
\newblock


\bibitem[Li et~al\mbox{.}(2022)]%
        {li2022adver}
\bibfield{author}{\bibinfo{person}{Xiaochen Li}, \bibinfo{person}{Jian Liang},
  \bibinfo{person}{Xialong Liu}, {and} \bibinfo{person}{Yu Zhang}.}
  \bibinfo{year}{2022}\natexlab{}.
\newblock \showarticletitle{Adversarial Filtering Modeling on Long-term User
  Behavior Sequences for Click-Through Rate Prediction}. In
  \bibinfo{booktitle}{\emph{{SIGIR}}}. \bibinfo{publisher}{{ACM}},
  \bibinfo{pages}{1969--1973}.
\newblock


\bibitem[Li et~al\mbox{.}(2021a)]%
        {yang2021light}
\bibfield{author}{\bibinfo{person}{Yang Li}, \bibinfo{person}{Tong Chen},
  \bibinfo{person}{Peng{-}Fei Zhang}, {and} \bibinfo{person}{Hongzhi Yin}.}
  \bibinfo{year}{2021}\natexlab{a}.
\newblock \showarticletitle{Lightweight Self-Attentive Sequential
  Recommendation}. In \bibinfo{booktitle}{\emph{{CIKM}}}.
  \bibinfo{publisher}{{ACM}}, \bibinfo{pages}{967--977}.
\newblock


\bibitem[Lian et~al\mbox{.}(2018)]%
        {xdeepfm}
\bibfield{author}{\bibinfo{person}{Jianxun Lian}, \bibinfo{person}{Xiaohuan
  Zhou}, \bibinfo{person}{Fuzheng Zhang}, \bibinfo{person}{Zhongxia Chen},
  \bibinfo{person}{Xing Xie}, {and} \bibinfo{person}{Guangzhong Sun}.}
  \bibinfo{year}{2018}\natexlab{}.
\newblock \showarticletitle{xDeepFM: Combining Explicit and Implicit Feature
  Interactions for Recommender Systems}. In \bibinfo{booktitle}{\emph{{KDD}}}.
  \bibinfo{publisher}{{ACM}}, \bibinfo{pages}{1754--1763}.
\newblock


\bibitem[Liu et~al\mbox{.}(2021)]%
        {chang2021non}
\bibfield{author}{\bibinfo{person}{Chang Liu}, \bibinfo{person}{Xiaoguang Li},
  \bibinfo{person}{Guohao Cai}, \bibinfo{person}{Zhenhua Dong},
  \bibinfo{person}{Hong Zhu}, {and} \bibinfo{person}{Lifeng Shang}.}
  \bibinfo{year}{2021}\natexlab{}.
\newblock \showarticletitle{Non-invasive Self-attention for Side Information
  Fusion in Sequential Recommendation}.
\newblock \bibinfo{journal}{\emph{CoRR}}  \bibinfo{volume}{abs/2103.03578}
  (\bibinfo{year}{2021}).
\newblock


\bibitem[Lu et~al\mbox{.}(2020)]%
        {wutong2020dual}
\bibfield{author}{\bibinfo{person}{Wantong Lu}, \bibinfo{person}{Yantao Yu},
  \bibinfo{person}{Yongzhe Chang}, \bibinfo{person}{Zhen Wang},
  \bibinfo{person}{Chenhui Li}, {and} \bibinfo{person}{Bo Yuan}.}
  \bibinfo{year}{2020}\natexlab{}.
\newblock \showarticletitle{A Dual Input-aware Factorization Machine for {CTR}
  Prediction}. In \bibinfo{booktitle}{\emph{{IJCAI}}}.
  \bibinfo{publisher}{ijcai.org}, \bibinfo{pages}{3139--3145}.
\newblock


\bibitem[Pi et~al\mbox{.}(2019)]%
        {mimn}
\bibfield{author}{\bibinfo{person}{Qi Pi}, \bibinfo{person}{Weijie Bian},
  \bibinfo{person}{Guorui Zhou}, \bibinfo{person}{Xiaoqiang Zhu}, {and}
  \bibinfo{person}{Kun Gai}.} \bibinfo{year}{2019}\natexlab{}.
\newblock \showarticletitle{Practice on Long Sequential User Behavior Modeling
  for Click-Through Rate Prediction}. In \bibinfo{booktitle}{\emph{{KDD}}}.
  \bibinfo{publisher}{{ACM}}, \bibinfo{pages}{2671--2679}.
\newblock


\bibitem[Pi et~al\mbox{.}(2020)]%
        {sim}
\bibfield{author}{\bibinfo{person}{Qi Pi}, \bibinfo{person}{Guorui Zhou},
  \bibinfo{person}{Yujing Zhang}, \bibinfo{person}{Zhe Wang},
  \bibinfo{person}{Lejian Ren}, \bibinfo{person}{Ying Fan},
  \bibinfo{person}{Xiaoqiang Zhu}, {and} \bibinfo{person}{Kun Gai}.}
  \bibinfo{year}{2020}\natexlab{}.
\newblock \showarticletitle{Search-based User Interest Modeling with Lifelong
  Sequential Behavior Data for Click-Through Rate Prediction}. In
  \bibinfo{booktitle}{\emph{{CIKM}}}. \bibinfo{publisher}{{ACM}},
  \bibinfo{pages}{2685--2692}.
\newblock


\bibitem[Qin et~al\mbox{.}(2021)]%
        {clea}
\bibfield{author}{\bibinfo{person}{Yuqi Qin}, \bibinfo{person}{Pengfei Wang},
  {and} \bibinfo{person}{Chenliang Li}.} \bibinfo{year}{2021}\natexlab{}.
\newblock \showarticletitle{The World is Binary: Contrastive Learning for
  Denoising Next Basket Recommendation}. In
  \bibinfo{booktitle}{\emph{{SIGIR}}}. \bibinfo{pages}{859--868}.
\newblock


\bibitem[Rashed et~al\mbox{.}(2022)]%
        {context2022ahmed}
\bibfield{author}{\bibinfo{person}{Ahmed Rashed}, \bibinfo{person}{Shereen
  Elsayed}, {and} \bibinfo{person}{Lars Schmidt{-}Thieme}.}
  \bibinfo{year}{2022}\natexlab{}.
\newblock \showarticletitle{Context and Attribute-Aware Sequential
  Recommendation via Cross-Attention}. In \bibinfo{booktitle}{\emph{RecSys}}.
  \bibinfo{publisher}{{ACM}}, \bibinfo{pages}{71--80}.
\newblock


\bibitem[Ren et~al\mbox{.}(2019)]%
        {hpmn}
\bibfield{author}{\bibinfo{person}{Kan Ren}, \bibinfo{person}{Jiarui Qin},
  \bibinfo{person}{Yuchen Fang}, \bibinfo{person}{Weinan Zhang},
  \bibinfo{person}{Lei Zheng}, \bibinfo{person}{Weijie Bian},
  \bibinfo{person}{Guorui Zhou}, \bibinfo{person}{Jian Xu},
  \bibinfo{person}{Yong Yu}, \bibinfo{person}{Xiaoqiang Zhu}, {and}
  \bibinfo{person}{Kun Gai}.} \bibinfo{year}{2019}\natexlab{}.
\newblock \showarticletitle{Lifelong Sequential Modeling with Personalized
  Memorization for User Response Prediction}. In
  \bibinfo{booktitle}{\emph{{SIGIR}}}. \bibinfo{publisher}{{ACM}},
  \bibinfo{pages}{565--574}.
\newblock


\bibitem[Ren et~al\mbox{.}(2020)]%
        {ren2020sequential}
\bibfield{author}{\bibinfo{person}{Ruiyang Ren}, \bibinfo{person}{Zhaoyang
  Liu}, \bibinfo{person}{Yaliang Li}, \bibinfo{person}{Wayne~Xin Zhao},
  \bibinfo{person}{Hui Wang}, \bibinfo{person}{Bolin Ding}, {and}
  \bibinfo{person}{Ji{-}Rong Wen}.} \bibinfo{year}{2020}\natexlab{}.
\newblock \showarticletitle{Sequential Recommendation with Self-Attentive
  Multi-Adversarial Network}. In \bibinfo{booktitle}{\emph{{SIGIR}}}.
  \bibinfo{publisher}{{ACM}}, \bibinfo{pages}{89--98}.
\newblock


\bibitem[Rendle(2010)]%
        {fm}
\bibfield{author}{\bibinfo{person}{Steffen Rendle}.}
  \bibinfo{year}{2010}\natexlab{}.
\newblock \showarticletitle{Factorization Machines}. In
  \bibinfo{booktitle}{\emph{{ICDM}}}. \bibinfo{publisher}{{IEEE} Computer
  Society}, \bibinfo{pages}{995--1000}.
\newblock


\bibitem[Singer et~al\mbox{.}(2022)]%
        {uriel2022sequential}
\bibfield{author}{\bibinfo{person}{Uriel Singer}, \bibinfo{person}{Haggai
  Roitman}, \bibinfo{person}{Yotam Eshel}, \bibinfo{person}{Alexander Nus},
  \bibinfo{person}{Ido Guy}, \bibinfo{person}{Or Levi}, \bibinfo{person}{Idan
  Hasson}, {and} \bibinfo{person}{Eliyahu Kiperwasser}.}
  \bibinfo{year}{2022}\natexlab{}.
\newblock \showarticletitle{Sequential Modeling with Multiple Attributes for
  Watchlist Recommendation in E-Commerce}. In
  \bibinfo{booktitle}{\emph{{WSDM}}}. \bibinfo{publisher}{{ACM}},
  \bibinfo{pages}{937--946}.
\newblock


\bibitem[Sun et~al\mbox{.}(2019)]%
        {bert4rec}
\bibfield{author}{\bibinfo{person}{Fei Sun}, \bibinfo{person}{Jun Liu},
  \bibinfo{person}{Jian Wu}, \bibinfo{person}{Changhua Pei},
  \bibinfo{person}{Xiao Lin}, \bibinfo{person}{Wenwu Ou}, {and}
  \bibinfo{person}{Peng Jiang}.} \bibinfo{year}{2019}\natexlab{}.
\newblock \showarticletitle{BERT4Rec: Sequential Recommendation with
  Bidirectional Encoder Representations from Transformer}. In
  \bibinfo{booktitle}{\emph{{CIKM}}}. \bibinfo{pages}{1441--1450}.
\newblock


\bibitem[Tian et~al\mbox{.}(2023)]%
        {eulernet}
\bibfield{author}{\bibinfo{person}{Zhen Tian}, \bibinfo{person}{Ting Bai},
  \bibinfo{person}{Wayne~Xin Zhao}, \bibinfo{person}{Ji{-}Rong Wen}, {and}
  \bibinfo{person}{Zhao Cao}.} \bibinfo{year}{2023}\natexlab{}.
\newblock \showarticletitle{EulerNet: Adaptive Feature Interaction Learning via
  Euler's Formula for {CTR} Prediction}. In
  \bibinfo{booktitle}{\emph{{SIGIR}}}. \bibinfo{publisher}{{ACM}},
  \bibinfo{pages}{1376--1385}.
\newblock


\bibitem[Tong et~al\mbox{.}(2021)]%
        {rap}
\bibfield{author}{\bibinfo{person}{Xiaohai Tong}, \bibinfo{person}{Pengfei
  Wang}, \bibinfo{person}{Chenliang Li}, \bibinfo{person}{Long Xia}, {and}
  \bibinfo{person}{Shaozhang Niu}.} \bibinfo{year}{2021}\natexlab{}.
\newblock \showarticletitle{Pattern-enhanced Contrastive Policy Learning
  Network for Sequential Recommendation}. In
  \bibinfo{booktitle}{\emph{{IJCAI}}}. \bibinfo{publisher}{ijcai.org},
  \bibinfo{pages}{1593--1599}.
\newblock


\bibitem[Velickovic et~al\mbox{.}(2018)]%
        {gat}
\bibfield{author}{\bibinfo{person}{Petar Velickovic}, \bibinfo{person}{Guillem
  Cucurull}, \bibinfo{person}{Arantxa Casanova}, \bibinfo{person}{Adriana
  Romero}, \bibinfo{person}{Pietro Li{\`{o}}}, {and} \bibinfo{person}{Yoshua
  Bengio}.} \bibinfo{year}{2018}\natexlab{}.
\newblock \showarticletitle{Graph Attention Networks}. In
  \bibinfo{booktitle}{\emph{{ICLR}}}. \bibinfo{publisher}{OpenReview.net}.
\newblock


\bibitem[Villegas et~al\mbox{.}(2018)]%
        {norha2018chara}
\bibfield{author}{\bibinfo{person}{Norha~M. Villegas},
  \bibinfo{person}{Cristian S{\'{a}}nchez}, \bibinfo{person}{Javier
  D{\'{\i}}az{-}Cely}, {and} \bibinfo{person}{Gabriel Tamura}.}
  \bibinfo{year}{2018}\natexlab{}.
\newblock \showarticletitle{Characterizing context-aware recommender systems:
  {A} systematic literature review}.
\newblock \bibinfo{journal}{\emph{Knowl. Based Syst.}}  \bibinfo{volume}{140}
  (\bibinfo{year}{2018}), \bibinfo{pages}{173--200}.
\newblock


\bibitem[Wang et~al\mbox{.}(2021)]%
        {dcnv2}
\bibfield{author}{\bibinfo{person}{Ruoxi Wang}, \bibinfo{person}{Rakesh
  Shivanna}, \bibinfo{person}{Derek~Zhiyuan Cheng}, \bibinfo{person}{Sagar
  Jain}, \bibinfo{person}{Dong Lin}, \bibinfo{person}{Lichan Hong}, {and}
  \bibinfo{person}{Ed~H. Chi}.} \bibinfo{year}{2021}\natexlab{}.
\newblock \showarticletitle{{DCN} {V2:} Improved Deep {\&} Cross Network and
  Practical Lessons for Web-scale Learning to Rank Systems}. In
  \bibinfo{booktitle}{\emph{{WWW}}}. \bibinfo{publisher}{{ACM} / {IW3C2}},
  \bibinfo{pages}{1785--1797}.
\newblock


\bibitem[Wu et~al\mbox{.}(2019)]%
        {SRGNN}
\bibfield{author}{\bibinfo{person}{Shu Wu}, \bibinfo{person}{Yuyuan Tang},
  \bibinfo{person}{Yanqiao Zhu}, \bibinfo{person}{Liang Wang},
  \bibinfo{person}{Xing Xie}, {and} \bibinfo{person}{Tieniu Tan}.}
  \bibinfo{year}{2019}\natexlab{}.
\newblock \showarticletitle{Session-Based Recommendation with Graph Neural
  Networks}. In \bibinfo{booktitle}{\emph{{AAAI}}}. \bibinfo{publisher}{{AAAI}
  Press}, \bibinfo{pages}{346--353}.
\newblock


\bibitem[Wu et~al\mbox{.}(2021)]%
        {limarec}
\bibfield{author}{\bibinfo{person}{Yongji Wu}, \bibinfo{person}{Lu Yin},
  \bibinfo{person}{Defu Lian}, \bibinfo{person}{Mingyang Yin},
  \bibinfo{person}{Neil~Zhenqiang Gong}, \bibinfo{person}{Jingren Zhou}, {and}
  \bibinfo{person}{Hongxia Yang}.} \bibinfo{year}{2021}\natexlab{}.
\newblock \showarticletitle{Rethinking Lifelong Sequential Recommendation with
  Incremental Multi-Interest Attention}.
\newblock \bibinfo{journal}{\emph{CoRR}}  \bibinfo{volume}{abs/2105.14060}
  (\bibinfo{year}{2021}).
\newblock


\bibitem[Zhang et~al\mbox{.}(2019)]%
        {zhang2019feature}
\bibfield{author}{\bibinfo{person}{Tingting Zhang}, \bibinfo{person}{Pengpeng
  Zhao}, \bibinfo{person}{Yanchi Liu}, \bibinfo{person}{Victor~S. Sheng},
  \bibinfo{person}{Jiajie Xu}, \bibinfo{person}{Deqing Wang},
  \bibinfo{person}{Guanfeng Liu}, {and} \bibinfo{person}{Xiaofang Zhou}.}
  \bibinfo{year}{2019}\natexlab{}.
\newblock \showarticletitle{Feature-level Deeper Self-Attention Network for
  Sequential Recommendation}. In \bibinfo{booktitle}{\emph{{IJCAI}}}.
  \bibinfo{publisher}{ijcai.org}, \bibinfo{pages}{4320--4326}.
\newblock


\bibitem[Zhang et~al\mbox{.}(2022)]%
        {ubcs}
\bibfield{author}{\bibinfo{person}{Yuren Zhang}, \bibinfo{person}{Enhong Chen},
  \bibinfo{person}{Binbin Jin}, \bibinfo{person}{Hao Wang},
  \bibinfo{person}{Min Hou}, \bibinfo{person}{Wei Huang}, {and}
  \bibinfo{person}{Runlong Yu}.} \bibinfo{year}{2022}\natexlab{}.
\newblock \showarticletitle{Clustering based Behavior Sampling with Long
  Sequential Data for {CTR} Prediction}. In
  \bibinfo{booktitle}{\emph{{SIGIR}}}. \bibinfo{publisher}{{ACM}},
  \bibinfo{pages}{2195--2200}.
\newblock


\bibitem[Zhou et~al\mbox{.}(2019)]%
        {dien}
\bibfield{author}{\bibinfo{person}{Guorui Zhou}, \bibinfo{person}{Na Mou},
  \bibinfo{person}{Ying Fan}, \bibinfo{person}{Qi Pi}, \bibinfo{person}{Weijie
  Bian}, \bibinfo{person}{Chang Zhou}, \bibinfo{person}{Xiaoqiang Zhu}, {and}
  \bibinfo{person}{Kun Gai}.} \bibinfo{year}{2019}\natexlab{}.
\newblock \showarticletitle{Deep Interest Evolution Network for Click-Through
  Rate Prediction}. In \bibinfo{booktitle}{\emph{{AAAI}}}.
  \bibinfo{publisher}{{AAAI} Press}, \bibinfo{pages}{5941--5948}.
\newblock


\bibitem[Zhou et~al\mbox{.}(2018)]%
        {din}
\bibfield{author}{\bibinfo{person}{Guorui Zhou}, \bibinfo{person}{Xiaoqiang
  Zhu}, \bibinfo{person}{Chengru Song}, \bibinfo{person}{Ying Fan},
  \bibinfo{person}{Han Zhu}, \bibinfo{person}{Xiao Ma},
  \bibinfo{person}{Yanghui Yan}, \bibinfo{person}{Junqi Jin},
  \bibinfo{person}{Han Li}, {and} \bibinfo{person}{Kun Gai}.}
  \bibinfo{year}{2018}\natexlab{}.
\newblock \showarticletitle{Deep Interest Network for Click-Through Rate
  Prediction}. In \bibinfo{booktitle}{\emph{{KDD}}}.
  \bibinfo{publisher}{{ACM}}, \bibinfo{pages}{1059--1068}.
\newblock


\bibitem[Zhuang et~al\mbox{.}(2021)]%
        {transferlearning}
\bibfield{author}{\bibinfo{person}{Fuzhen Zhuang}, \bibinfo{person}{Zhiyuan
  Qi}, \bibinfo{person}{Keyu Duan}, \bibinfo{person}{Dongbo Xi},
  \bibinfo{person}{Yongchun Zhu}, \bibinfo{person}{Hengshu Zhu},
  \bibinfo{person}{Hui Xiong}, {and} \bibinfo{person}{Qing He}.}
  \bibinfo{year}{2021}\natexlab{}.
\newblock \showarticletitle{A Comprehensive Survey on Transfer Learning}.
\newblock \bibinfo{journal}{\emph{Proc. {IEEE}}} \bibinfo{volume}{109},
  \bibinfo{number}{1} (\bibinfo{year}{2021}), \bibinfo{pages}{43--76}.
\newblock


\end{thebibliography}
